# Hopf solitons in helical and conical backgrounds of chiral magnetic solids

Robert Voinescu[1#], Jung-Shen B. Tai (戴榮身)[1#] and Ivan I. Smalyukh[1,2,3*]

[1]Department of Physics, University of Colorado, Boulder, CO 80309, USA
[2]Materials Science and Engineering Program, Soft Materials Research Center and Department of Electrical, Computer and Energy Engineering, University of Colorado, Boulder, CO 80309, USA
[3]Renewable and Sustainable Energy Institute, National Renewable Energy Laboratory and University of Colorado, Boulder, CO 80309, USA

[#]These authors contributed equally to this study
[*]Corresponding author: ivan.smalyukh@colorado.edu

*Abstract: Three-dimensional topological solitons attract a great deal of interest in fields ranging from particle physics to cosmology but remain experimentally elusive in solid-state magnets. Here we numerically predict magnetic heliknotons, an embodiment of such nonzero-Hopf-index solitons localized in all spatial dimensions while embedded in a helical or conical background of chiral magnets. We describe conditions under which heliknotons emerge as metastable or ground-state localized nonsingular structures with fascinating knots of magnetization field in widely studied materials. We demonstrate magnetic control of three-dimensional spatial positions of such solitons, as well as show how they interact to form molecule-like clusters and possibly even crystalline phases comprising three-dimensional lattices of such solitons with both orientational and positional order. Finally, we discuss both fundamental importance and potential technological utility of magnetic heliknotons.*

While one-dimensional topological solitons, like magnetic Neel and Bloch domain walls $\pi_1(\mathbb{S}^1) = \mathbb{Z}$, have been studied for nearly a century [1–3], their higher-dimensional analogs remain relatively elusive, consistent with the Derrick-Hobart theorem [4] applied to systems where solitons are unstable against rescaling perturbations. Evading predictions of the theorem, diverse embodiments of such high-dimensional solitons were found in nuclear physics, superconductors, liquid crystals, magnets, optics and so on [5–20], with sources of stability ranging from high-order-derivative terms in the Skyrme model [21,22] and superconductors [23], to chiral terms in condensed matter [3,14–16]. In solid-state non-centrosymmetric magnets, the two-dimensional $\pi_2(\mathbb{S}^2) = \mathbb{Z}$ solitons, often called "baby skyrmions" to denote that they are lower-dimensional analogs of Skyrme's $\pi_3(\mathbb{S}^3) = \mathbb{Z}$ nuclear physics counterparts [3], are nowadays a major theme of fundamental and spintronics-inspired applied research [24,25]. However, the three dimensional (3D) Hopf solitons, $\pi_3(\mathbb{S}^2) = \mathbb{Z}$ topological solitons localized in all three spatial dimensions [5,6], remain experimentally elusive in magnetic solids. Predictions of such solitons [8–10,26] embedded within the ferromagnetic background remain to be experimentally tested. On the other hand, individual and 3D crystalline lattices of Hopf solitons (called "heliknotons") were recently demonstrated in a helical background of cholesteric liquid crystals [17].

In this work we predict heliknotons embedded in the helical and conical backgrounds of bulk chiral magnets. These heliknotons display Hopf-fibration-like linking of preimages in the magnetization field $\boldsymbol{m}(\boldsymbol{r})$ and singular vortex lines forming links and knots in the non-polar immaterial helical wavevector field $\boldsymbol{q}(\boldsymbol{r})$. We derive structural phase diagrams with (meta)stability of heliknotons by comparing their free energy to those of topologically trivial helical and conical states. We probe how stability of heliknotons is further controlled by experimentally accessible applied magnetic fields and magnetocrystalline anisotropies, which can arise due to crystal symmetry, mechanical stress or lattice mismatches [27–29]. We show that the position and orientation of magnetic heliknotons can be effectively controlled in 3D and that emergent magnetic field lines also form Hopf-fibration-like structures. Numerically simulated Lorentz Transmission

Electron Microscopy (LTEM) images have characteristic features that will guide experimental discovery of these topological solitons. Lastly, we study heliknoton interactions and oligomeric self-organizations, suggesting the possibility of 3D crystalline phases.

We use the standard micromagnetic Hamiltonian of a chiral magnet with the energy density $w = w_\text{i} + w_\text{Z} + w_\text{a}$, where

$$w_\text{i} = \left(\frac{J}{2}(\nabla \boldsymbol{m})^2 + D\boldsymbol{m} \cdot (\nabla \times \boldsymbol{m})\right), \quad (1)$$

$$w_\text{Z} = -\mu_0 M_s \boldsymbol{m} \cdot \boldsymbol{H}. \quad (2)$$

$w_\text{i}$ contains the Heisenberg exchange and Dzyaloshinskii-Moriya interaction terms with corresponding constants $J$ and $D$ defining the helical wavelength $\lambda = 2\pi(J/D)$. $w_\text{Z}$ is the Zeeman coupling term, where $\boldsymbol{H}$ is the applied field, $M_s$ is the saturation magnetization, and $\mu_0$ is the vacuum permeability. The magnetocrystalline anisotropy term is $w_\text{a} = -K_\text{u}(\boldsymbol{m} \cdot \hat{k})^2$ for uniaxial anisotropy and $w_\text{a} = -K_\text{c}(m_x^4 + m_y^4 + m_z^4)$ for cubic anisotropy, where $K_\text{u}$ and $\hat{k}$ are the uniaxial anisotropy strength and axis, and $K_\text{c}$ is the cubic anisotropy strength. To make our findings applicable to a broad range of materials [30], we use dimensionless fields and anisotropy strengths $\widetilde{H} = H/H_D$ and $\widetilde{K}_\text{u(c)} = K_\text{u(c)}/\mu_0 M_s H_D$, where $H_D = D^2/\mu_0 M_s J$ is the critical field along the helical axis that fully unwinds the helical state [30].

Adopting the field configuration of liquid-crystal heliknotons in Ref. [17] as the initial condition of $\boldsymbol{m}(\boldsymbol{r})$, we minimize free energy and find that the individual 3D-localized magnetic heliknotons in the bulk helical background at no fields or anisotropies [Fig. 1(a-d)] as metastable states with energy $E_0 = 8.58\,J\lambda$ when taking the helical background state as the reference. Preimages of constant $\boldsymbol{m}(\boldsymbol{r})$ corresponding to $\mathbb{S}^2$-points are closed-loops interlinking once with other individual preimages. This geometric analysis allows the assignment of the Hopf index $Q =$ 1 to the heliknoton, which is consistent with the numerically calculated $Q$ up to numerical error [8,35]. The spatial extent of a heliknoton can be visualized by the isosurface of a small deviation of $\boldsymbol{q}(\boldsymbol{r})$ from the uniform far-field $\boldsymbol{q}_0$ [Fig. 1(e)]. Preimages of $\mathbb{S}^2$-points of constant polar but different azimuthal angles form deformed tori nested around the preimages of north and south poles [Fig. 1(f)]. The two sets of nested tori are separated by the preimages of points on the equator of $\mathbb{S}^2$, representing the region occupied by the far-field. In a heliknoton embedded in a helical background, preimage tori corresponding to points of the same latitude on either hemisphere of $\mathbb{S}^2$ inter-transform by a $\pi$ rotation along $\boldsymbol{q}_0$ with respect to the geometric center of the heliknoton. This symmetry is broken when a magnetic field is applied along $\boldsymbol{q}_0$ and the helical background transitions into the conical state with a cone angle $\theta_c = \cos^{-1}\widetilde{H}$ [Fig. 1(a)]. As a result of such helical-to-conical transition in the far-field, two preimage tori of polar angles $\theta_1$ and $\theta_2$ ($\theta_1 < \theta_c < \theta_2 < \pi/2$) transition from being both coaxial with the north pole's preimage to forming a non-coaxial link of preimage tori, with the overall $\pi_3(\mathbb{S}^2)$ topology of $\boldsymbol{m}(\boldsymbol{r})$ [Fig. 1(g)]. Thus, heliknotons can exist in a conical field background of varying cone angle, though we could stabilize heliknotons only up to $\widetilde{H} \approx 0.2$. Beyond this field, the high energy cost of regions with $\boldsymbol{m}(\boldsymbol{r})$ antiparallel to $\boldsymbol{H}$ overcomes the topological barrier, transforming the $Q = 1$ heliknoton to the topologically trivial conical state through nucleation and propagation of singular defects (Bloch points) [35].

While heliknotons are fully nonsingular structures in $\boldsymbol{m}(\boldsymbol{r})$, nontrivial topology characterizes not only this material field. Singular vortex lines in non-polar $\boldsymbol{q}(\boldsymbol{r})$ form three mutually linked loops, different from the trefoil-knot vortices of liquid crystal heliknotons [17,30] [Figs. 1(h)]. We also calculate the emergent field $(B_{\text{em}})_i \equiv \hbar\varepsilon^{ijk}\boldsymbol{m} \cdot (\partial_j\boldsymbol{m} \times \partial_k\boldsymbol{m})/2$, a fictitious field describing the interaction between conduction electrons and the underlying spin texture related to many spintronic implications [36]. $\boldsymbol{B}_{\text{em}}$ correlates with the localized structure of $\boldsymbol{m}(\boldsymbol{r})$ and features closed-loop streamlines, with each pair of loops linked exactly once, once again resembling the Hopf fibration [Fig. 1(i)]. The topology of Hopf fibrations in $\boldsymbol{B}_{\text{em}}$-streamlines in both magnetic heliknotons and hopfions (embedded in a helical and uniform background, respectively) is a salient feature of their $\pi_3(\mathbb{S}^2)$ topology in $\boldsymbol{m}(\boldsymbol{r})$ [8]. To facilitate experimental discovery of such magnetic heliknotons, we numerically simulate their LTEM images using energy-minimizing $\boldsymbol{m}(\boldsymbol{r})$ [Fig. 1(j-l)], which significantly differ from those of other topological states found so far [19]. Recent advances in 3D imaging of $\boldsymbol{m}(\boldsymbol{r})$ could also assist in unambiguously showing their existence [37].

When a magnetic field $\boldsymbol{H}$ is applied perpendicular to $\boldsymbol{q}_0$, the helical state transforms from a harmonic modulation to distorted helicoids [38,39]. The energy of a heliknoton hosted within a distorted helicoidal background depends on the relative orientation of $\boldsymbol{H}$ and the orientation of a heliknoton defined by the magnetization $\boldsymbol{m}_\text{h}$ at its geometric center. At $\widetilde{H} = 0.05$, heliknotons remain stationary when $\boldsymbol{H}$ is either parallel or antiparallel to $\boldsymbol{m}_\text{h}$, while the energy difference between the former and the latter case is $\Delta E = 0.13 \, \lambda J$, $E_{\text{parallel}} = E_{\text{antiparallel}} + \Delta E$ [Fig. 2(a)]. With the heliknoton being a metastable excitation within the helicoidal background, $\boldsymbol{H}$ parallel (antiparallel) to $\boldsymbol{m}_\text{h}$ expands (contracts) [Fig. 2(a)] the spatial extent of $\boldsymbol{m}(\boldsymbol{r})$-distortions. A small variation of the angle between $\boldsymbol{H}$ and $\boldsymbol{m}_\text{h}$ drives the heliknoton away from the metastable $\boldsymbol{m}_\text{h} \parallel \boldsymbol{H}$ state, and the heliknoton undergoes a screw-like motion of correlated rotation and displacement along $\boldsymbol{q}_0$ in a sense consistent with the material chirality, eventually arriving at the antiparallelly-aligned state [Fig. 2(b)]. Figure 2(c) shows the heliknoton's energy difference versus orientation and the correlated vertical displacement during this motion. At fields as small as $\widetilde{H} = 0.05$, heliknotons can be perturbed out of metastability when the angle between $\boldsymbol{H}$ and $\boldsymbol{m}_\text{h}$ is as small as 0.01°. At larger fields, a larger angle between $H$ and $m_h$ is required to drive the screw motion due to the adaptive deformation of $m(r)$ and corresponding pinning of the heliknoton by $H$. For $H \perp m_h$ and rotating synchronously with the screw motion of heliknotons, such rotating-wave magnetic field can be used to control orientations and positions of heliknotons [30].

We explore heliknoton stability at applied fields and various magnetocrystalline anisotropies by comparing free energy to that of topologically trivial helical (distorted helicoidal) and conical states (Fig. 3). The fields are collinear with $m_h$ and take positive values when parallel to $m_h$. We consider $|\widetilde{H}| \leq 0.45$, beyond which significant stretching and distortion in the heliknotons take place. Three cases are considered: uniaxial easy-plane anisotropy with the hard-axis $\hat{k} \parallel q_0$ [$K_u < 0$, Fig. 3(a)], uniaxial easy-axis anisotropy with the easy-axis $\hat{k} \perp q_0$ and collinear with $H$ [$K_u > 0$, Fig. 3(b)], and cubic anisotropy [Fig. 3(c)]. For uniaxial anisotropies, heliknotons are always higher energy than the topologically trivial structures but persist as metastable states within a broad parameter range (labeled green in Fig. 3). Within metastability regions, these solitons are often geometrically deformed by fields and anisotropies (Fig. 4) [30]. Interestingly, this stretching preserves topology and Hopf index but sometimes alters the singular vortex lines in the immaterial field $q(r)$ (Fig. 4). The stretching occurs for both positive and negative $H$, though it is more prevalent for antiparallel $H$ and $m_h$, providing a means for the geometric control of heliknotons. In the cases of easy-plane and easy-axis uniaxial anisotropies, the conical state with helical axis colinear with $H$ is the lowest energy state at higher fields and the energy difference between heliknotons and conical states increases with $|\widetilde{H}|$, yielding the valley-like energy surfaces along the field axis [Figs. 3(a,b)]. At stronger anisotropy strengths, easy-plane anisotropy tends to restore the helical state [Fig. 3(a)] and easy-axis anisotropy tends to unwind the twisted structures [Fig. 3(b)], both tending to destabilize heliknotons (gray areas in Fig. 3). The metastability range of heliknotons against strong uniaxial anisotropies is extended with the applied field strength, particularly in the case of positive fields ($H \parallel m_h \parallel \hat{y}$). This is because regions with $m(r) \parallel H$ are anchored against the destabilizing uniaxial anisotropies, and the expansion induced by positive $H$ helps counteract the shrinking of heliknotons during destabilization [Fig. 2(a)]. For cubic anisotropy with $K_c < 0$ (hard axes for $\langle 100 \rangle$ directions), heliknotons have energy even lower than that of the conical and helical states in some parameter region and are the globally stable state (labeled purple in Fig. 3c). This is a result of delicate competition between different free energy terms. At strong cubic anisotropy and $K_c > 0$, heliknotons become unstable, but they are also metastable within a broad range of parameters which correspond to widely studied material systems like MnSi, FeGe, $Cu_2OSeO_3$, etc [18,27,40]. Interestingly, cubic anisotropy has also been reported to be critical for observing a novel solitonic state found at temperatures lower than that of the conventional A-phase of skyrmions [27,41]. Our findings indicate that chiral magnetic materials with cubic anisotropy and $K_c < 0$ are the best candidates for observing magnetic hopfions of heliknoton type introduced here (Fig. 3c).

Heliknotons interact by sharing perturbations in the helical or conical background around them and minimizing the overall free energy of multi-soliton configurations versus their relative 3D positions and orientations. Figures 5(a-b) show the positive (red) and negative (green) perturbations relative to the helical background energy density around an individual heliknoton in an isotropic chiral magnetic at no fields. Starting with two-heliknoton configurations with different orientations relative to the separation vector, heliknotons display anisotropic attractive interactions and can form three different dimer configurations [Fig. 5(c)] [30]. The energy per heliknoton for all dimer configurations is reduced as compared to that of an individual heliknoton $E_0$ as a result of sharing regions of high-energy $m(r)$-distortions. The energy differences between different dimer configurations can be as small as $\Delta E_{23} = 0.045\,J\lambda$ between dimer-2 and dimer-3, which, depending on material parameters, can be large ($\Delta E_{23}/k_BT \approx 10$ and $T = 200$ K for FeGe) or comparable ($\Delta E_{23}/k_BT \approx 0.7$ and $T = 25$ K for MnSi) to thermal energy [30]. With the formation of tetramer and octamer configurations, the free energy per heliknoton is further reduced. Within the heliknoton oligomer, the isosurfaces of perturbation in $\boldsymbol{q}(\boldsymbol{r})$ of individual heliknotons join into a single surface and the overall Hopf index becomes the sum of that of the solitonic constituents [Fig. 5(d-e)]. Thus, a heliknoton oligomer resembles a single high-charge heliknoton molecule, or, in a different analogy, a high-baryon-number nucleus [42]. The complex configuration of the stabilized octamer cannot be straightforwardly expected on the basis of dimer or tetramer configurations, suggesting that the emergent crystalline assemblies of heliknotons could be complex. A systematic study of all possible symmetries and lattice parameters, for different external fields and magneto-crystalline anisotropies, could potentially reveal the energy-minimizing assembly of various heliknoton crystals and potential solitonic condensed matter phases, again drawing an analogy to skyrmion crystals [5,6] and synthetic skyrmion crystals in certain superfluids [43], though such exploration is beyond the scope of this work.

To conclude, we have modeled metastable and stable heliknotons in the helical and conical backgrounds of bulk chiral magnetic materials. The demonstrated 3D-localization and magnetic spatial control of Hopf solitons within the helical background, combined with the conventional control of dynamical topological solitons by spin currents [24,26], may provide a versatile set of tools and properties needed for spintronics applications. Formation of clustered heliknoton oligomers suggests the possibility of using high-charge heliknotons as information carriers or as building blocks of topological phases. Our findings call for the experimental discoveries of magnetic heliknotons based on their unique LTEM textures or using other imaging techniques.

We acknowledge funding from the US Department of Energy, Office of Basic Energy Sciences, Division of Materials Sciences and Engineering, under Award ER46921, contract DE-SC0019293 with the University of Colorado at Boulder. This work utilized the RMACC Summit supercomputer, which is supported by the NSF (awards ACI-1532235 and ACI-1532236), the University of Colorado Boulder and Colorado State University.

**Figures:**

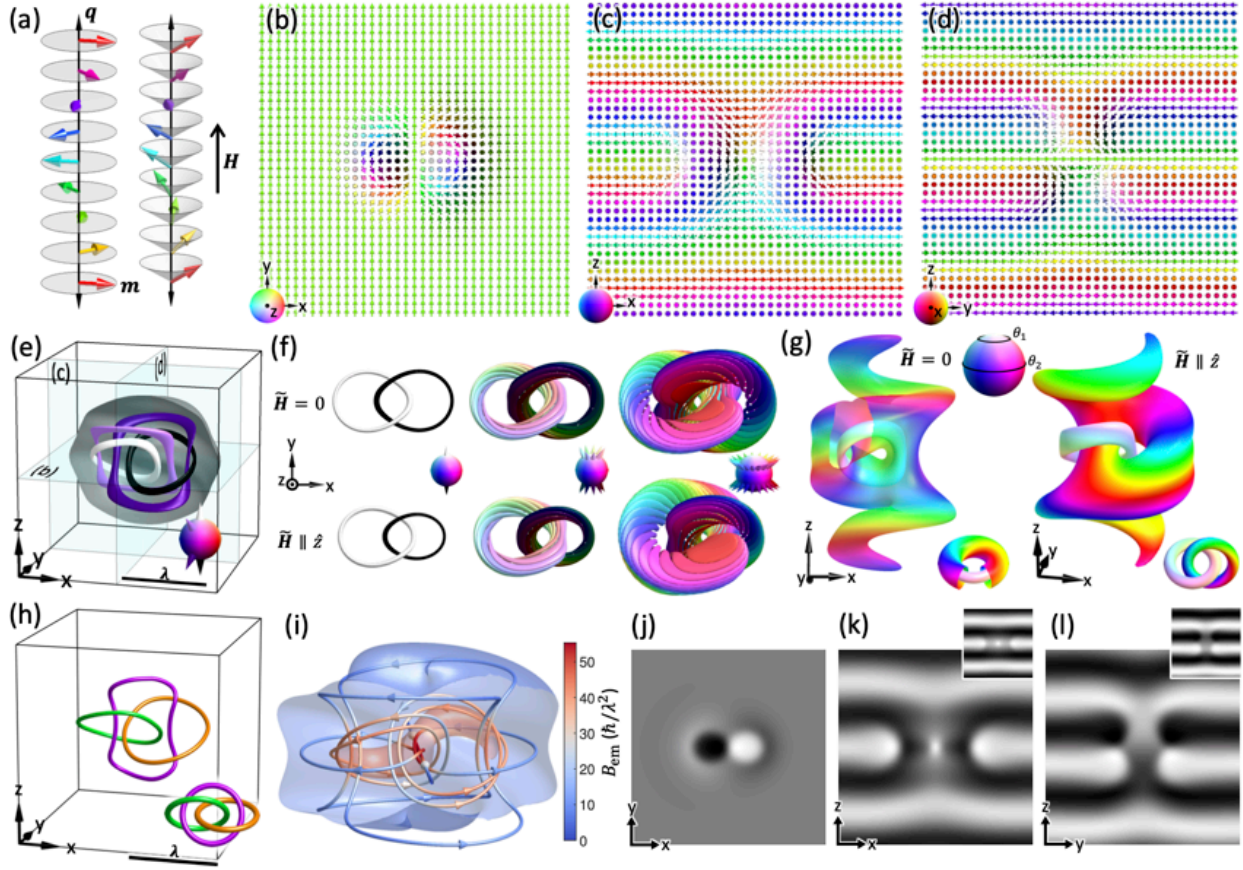

Fig. 1. Magnetic heliknoton. (a) Helical (left) and conical fields (right). (b,c,d) Simulated cross-sections of $m(r)$ of a heliknoton in a helical background. $m(r)$ is shown with arrows colored according to orientations on $\mathbb{S}^2$ (bottom-left insets). (e) Preimages in $m(r)$ of a heliknoton colored according to their orientations shown as cones on $\mathbb{S}^2$ (bottom-right inset). The gray isosurface (bisected for clarity) shows the region with small deviation of $q(r)$ from the background $q_0$. (f) Preimages of constant-polar-angle orientations as cones on $\mathbb{S}^2$ of heliknotons in a helical field (top, $\widetilde{H} = 0$) and in a conical field (bottom, $\widetilde{H} = 0.15\hat{z}$). (g) Constant-polar-angle surfaces at angles shown on $\mathbb{S}^2$ of heliknotons in a helical field (left, $\widetilde{H} = 0$) and in a conical field (right, $\widetilde{H} = 0.15\hat{z}$). Schematics of the nesting of tori surfaces are shown in the bottom-right insets. (h) Singular vortex lines in $q(r)$ forming three mutually linked rings (schematic in bottom-right inset). (i) Visualization of $B_{\mathrm{em}}$ in a magnetic heliknoton by the isosurfaces colored by magnitude and streamlines with cones. (j,k,l) Simulated LTEM images of a heliknoton in (b,c,d) for different directions. Top-right insets in (k) and (l) show the images with the characteristic contrast of a heliknoton reduced in thick samples.

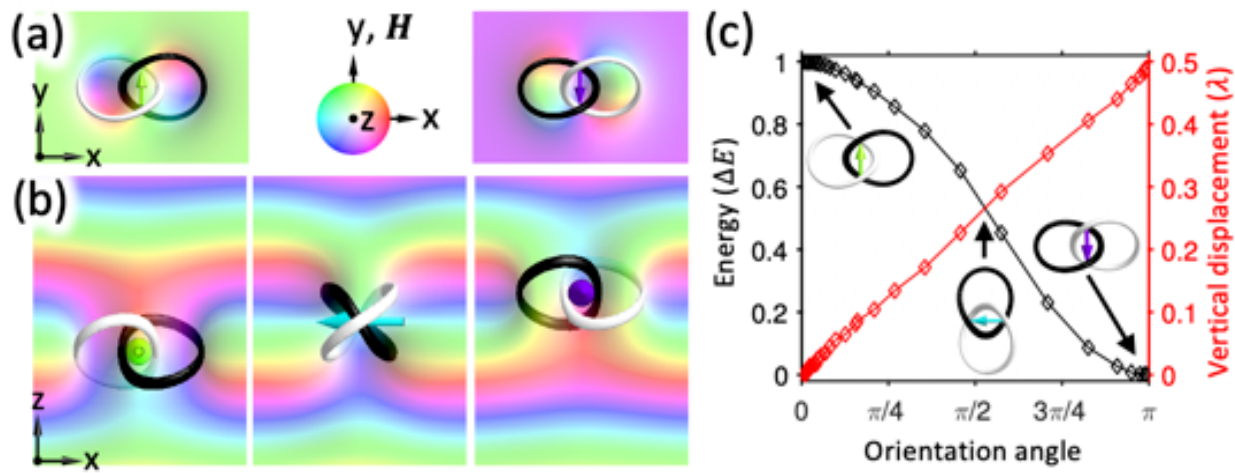

FIG. 2. Screw motions of heliknotons induced by external magnetic fields. (a) Heliknotons, visualized by preimages of poles, with orientations $m_h$ parallel (left) or antiparallel (right) to the applied field being the metastable and stable state, respectively. (b) Screw motion of a heliknoton at $\tilde{H} = 0.05\hat{y}$ relaxing from metastable parallel to stable antiparallel orientation. (c) The energy, orientation, and vertical displacement of a heliknoton relaxing from the metastable parallel configuration to the stable antiparallel configuration upon a perturbing magnetic field; the energy is normalized by $\Delta E = E_{\text{parallel}} - E_{\text{antiparallel}}$ and referenced by $E_{\text{antiparallel}}$.

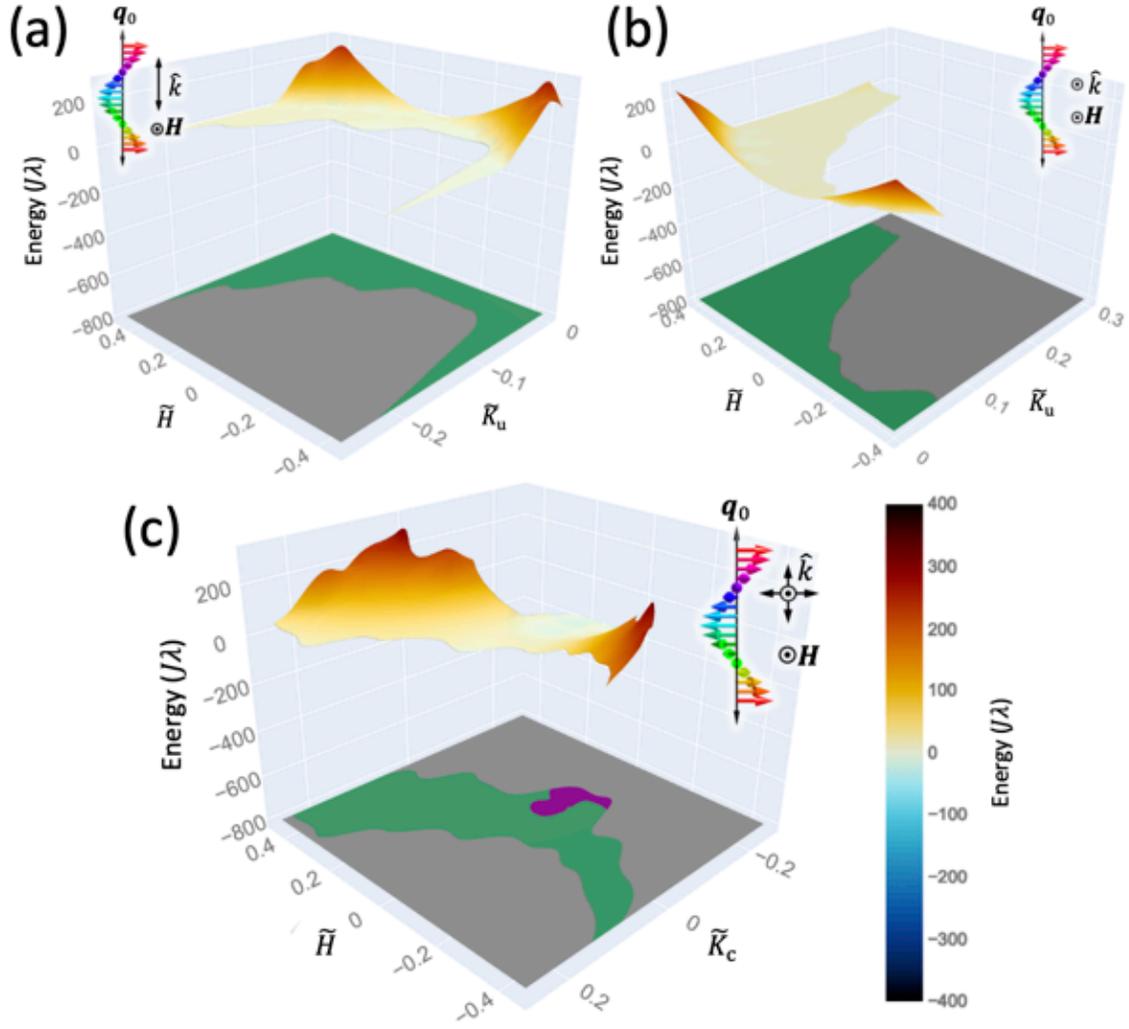

FIG. 3. Heliknoton stability. (a,b,c) Energy surfaces and stability diagrams of heliknotons in a chiral magnetic with (a) uniaxial easy-plane anisotropy, (b) uniaxial easy-axis anisotropy, and (c) cubic anisotropy. The energy is presented as the difference between the heliknoton energy and that of the minimum of helical and conical states. Colored regions indicate stable (purple), metastable (green), and unstable (gray) heliknotons. Relative orientations of $q_0$, $H$, and $\hat{k}$ are shown in the insets.

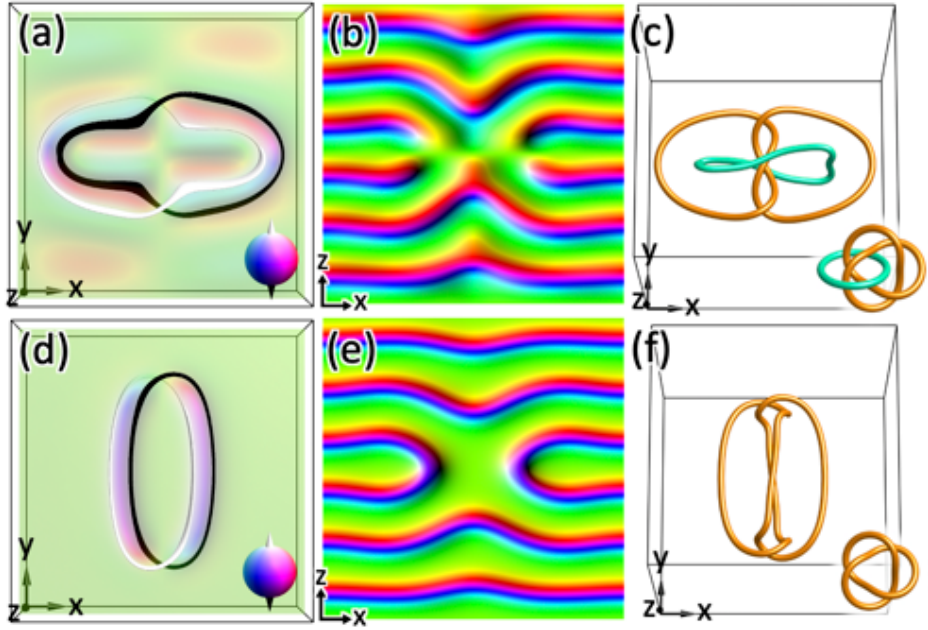

FIG. 4. Deformed heliknotons. (a,b,c) A heliknoton at $\tilde{\boldsymbol{H}} = 0.4\hat{y}$, shown using preimages of poles in $\mathbb{S}^2$ in (a) and $\boldsymbol{m}(\boldsymbol{r})$ colored according to orientations on $\mathbb{S}^2$ in mid-plane cross-sections in (a) and (b), and singular vortex loops in $\boldsymbol{q}(\boldsymbol{r})$ in (c) shown by colored tubes. (d,e,f) A heliknoton at $\tilde{\boldsymbol{H}} = 0.45\hat{y}$ and easy-axis uniaxial anisotropy $\tilde{K}_{\mathrm{u}} = 0.16$ along $\hat{y}$, shown by preimages of poles in (d) and $\boldsymbol{m}(\boldsymbol{r})$ colored according to orientations on $\mathbb{S}^2$ in mid-plane cross-sections in (d) and (e), and singular vortex lines in $\boldsymbol{q}(\boldsymbol{r})$ in (f).

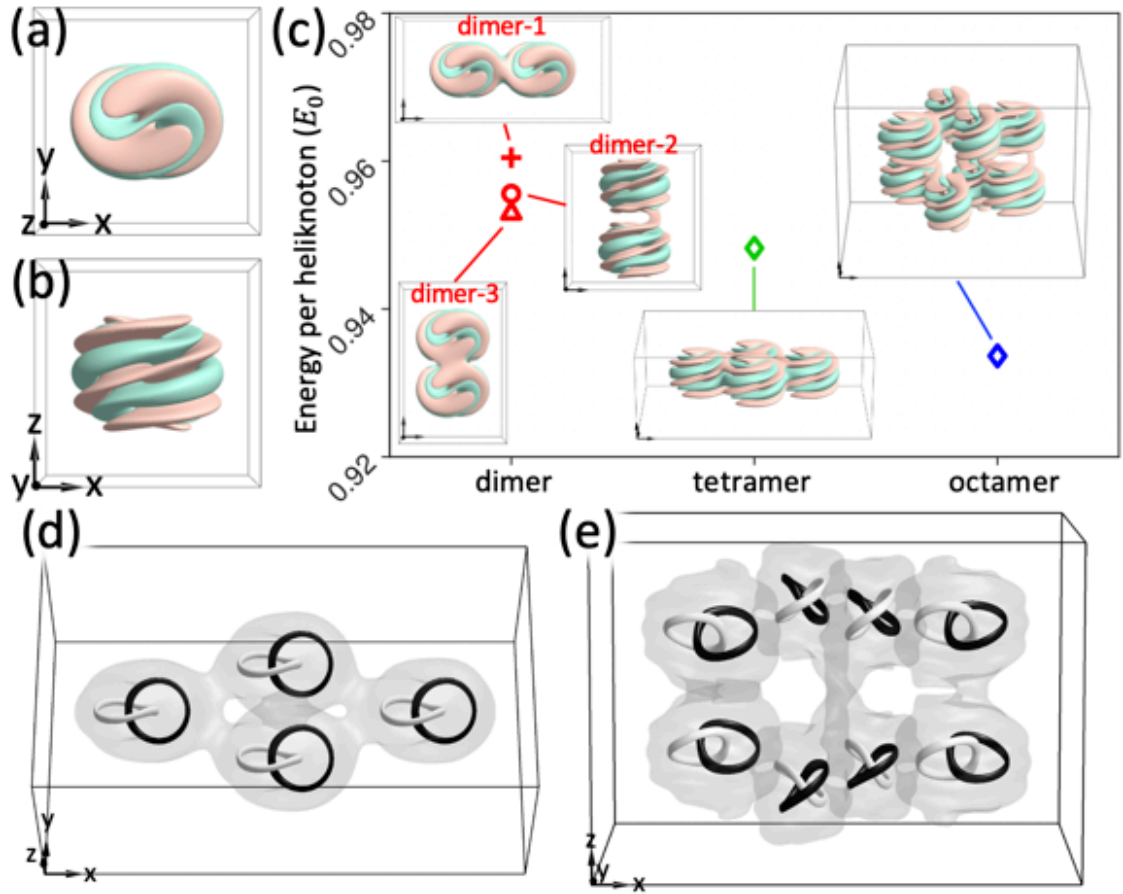

FIG. 5. Oligomeric assemblies of magnetic heliknotons. (a,b) Isosurfaces of positive (red) and negative (green) perturbations in the background energy density by an individual heliknoton. (c) Energy per soliton of dimers, tetramer and octamer, in units of $E_0$. Configurations of each oligomer are shown as insets. (d,e) Tetramer (d) and octamer (e) of heliknotons shown by preimages at poles and isosurfaces of small deviation of $\mathbf{q}(\mathbf{r})$ from the background.

which include Refs. [31-34].

# Supplemental Material for "Hopf solitons in helical and conical backgrounds of chiral magnetic solids"

Robert Voinescu[1#], Jung-Shen B. Tai (戴榮身)[1#] and Ivan I. Smalyukh[1,2,3*]

[1]Department of Physics, University of Colorado, Boulder, CO 80309, USA
[2]Materials Science and Engineering Program, Soft Materials Research Center and Department of Electrical, Computer and Energy Engineering, University of Colorado, Boulder, CO 80309, USA
[3]Renewable and Sustainable Energy Institute, National Renewable Energy Laboratory and University of Colorado, Boulder, CO 80309, USA
*Corresponding author: ivan.smalyukh@colorado.edu

## Supplementary Figures & Tables

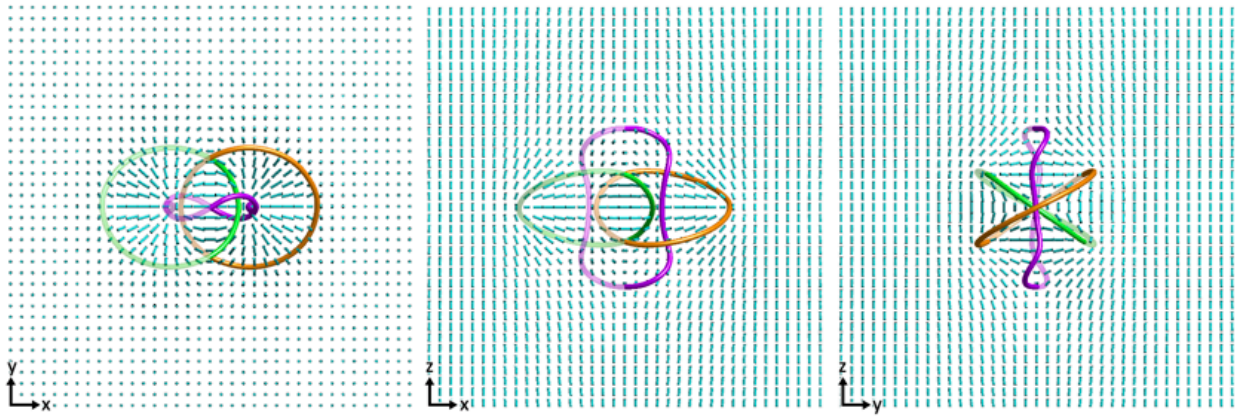

FIG. S1. Computer-simulated cross-sections of $q(r)$ of a heliknoton in different cross-sections shown with cylinders. Here the lengths of cylinders are scaled by the magnitude of the corresponding helical wavevector. The colored tubes represent half-integer singular vortex lines in in the immaterial field $q(r)$.

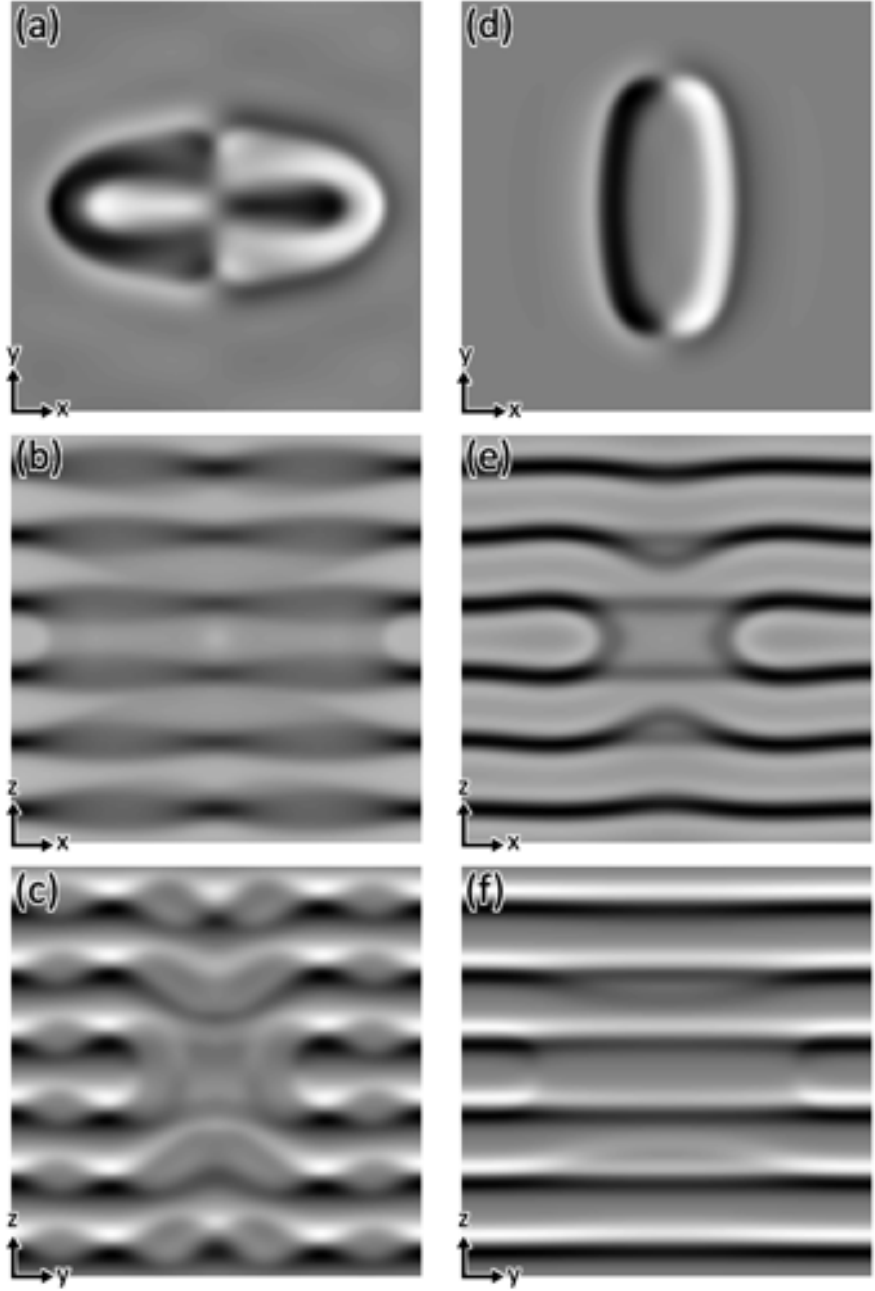

FIG. S2. Computer-simulated LTEM images of heliknotons at finite magnetic fields and anisotropies in Fig. 4. (a,b,c) LTEM images of a heliknoton at $\widetilde{\boldsymbol{H}} = 0.4\hat{y}$ in Fig. 4(a,b,c,d) imaged along different directions. (d,e,f) LTEM images of a heliknoton at $\widetilde{\boldsymbol{H}} = 0.45\hat{y}$ and easy-axis uniaxial anisotropy $\widetilde{K}_\mathrm{u} = 0.16$ along $\hat{y}$ in Fig. 4(e,f,g,h) imaged along different directions.

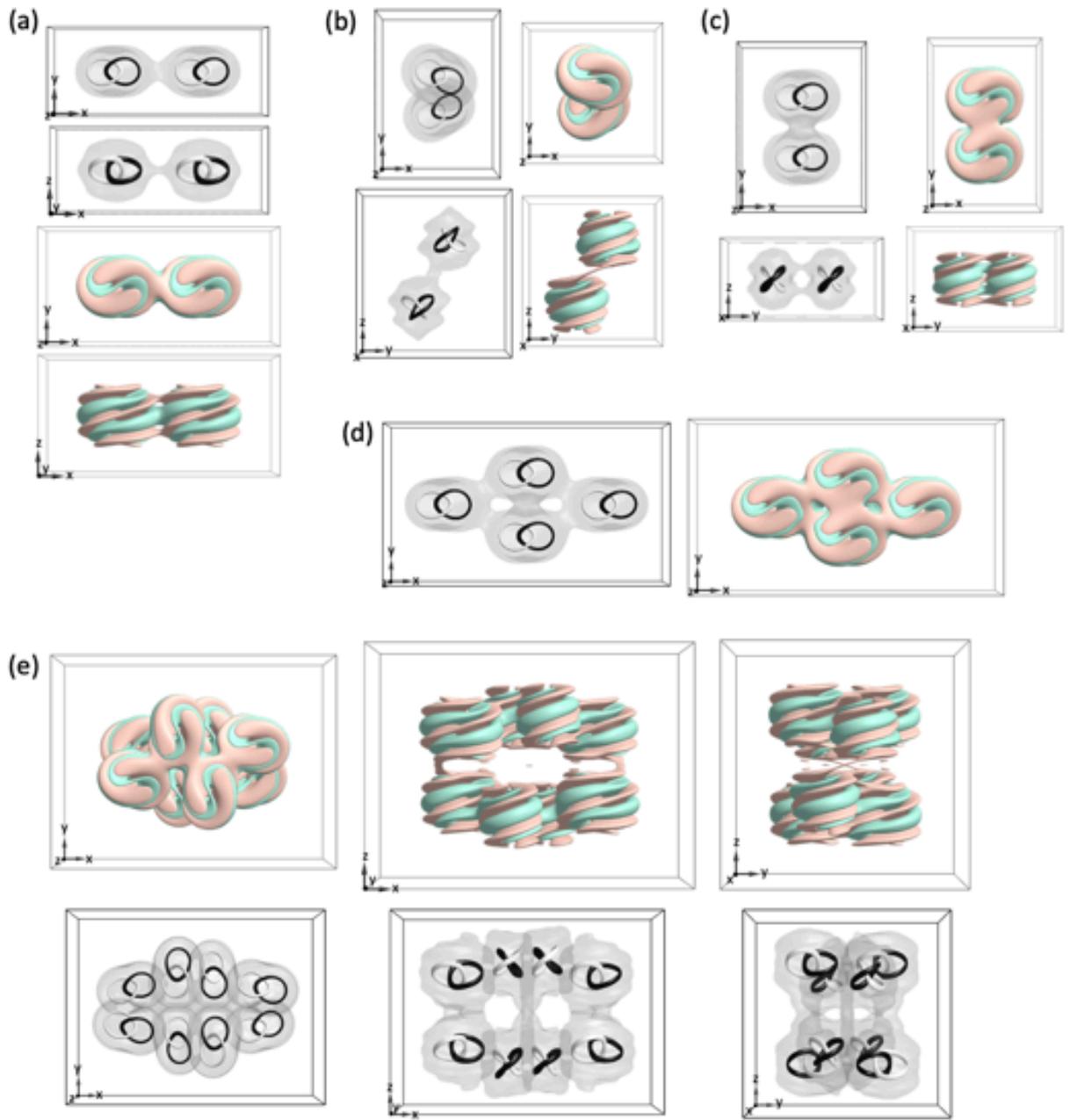

FIG. S3. Additional details for heliknoton oligomers. Isosurfaces of positive (red) and negative (green) perturbations in the background energy density and heliknoton oligomers shown by preimages at poles and isosurfaces of small deviation of $q(r)$ from the background for (a) dimer-1, (b) dimer-2, (c) dimer-3, (d) tetramer, and (e) octamer, respectively.

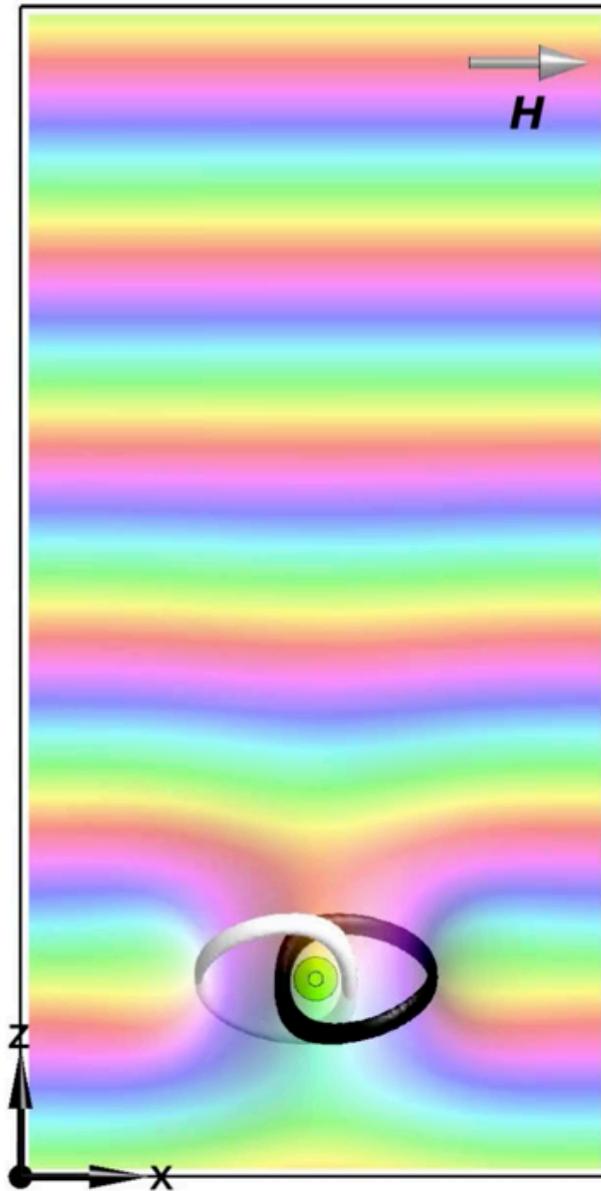

Movie S1. Screw motion of a heliknoton shown by preimages at poles and $m_h$, driven by a synchronously rotating $H$.

Table S1. Material parameters of FeGe [1] and MnSi [2].

| Material | $J$ | $D$ | $\lambda = 2\pi(\frac{J}{D})$ | $\mu_0 M_s$ |
|---|---|---|---|---|
| FeGe | 7.290 pJm$^{-1}$ | 0.567 mJm$^{-2}$ | 80.8 nm | 0.14 T |
| MnSi | 0.32 pJm$^{-1}$ | 0.115 mJm$^{-2}$ | 17.5 nm | 0.19 T |

## Supplementary Methods

The micromagnetic Hamiltonian of a bulk chiral magnet subject to external magnetic fields and magnetocrystalline anisotropy is given by

$$E = \int d^3\mathbf{r}\, w_i + w_z + w_a$$

$$= \int d^3\mathbf{r} \left[\frac{J}{2}(\nabla \mathbf{m})^2 + D\mathbf{m} \cdot (\nabla \times \mathbf{m}) - \mu_0 M_s \mathbf{m} \cdot \mathbf{H} + w_a\right] \quad (S1)$$

The energy density $w_i$ contains the Heisenberg exchange and Dzyaloshinskii-Moriya interaction terms with corresponding constants $J$ and $D$ defining the helical wavelength $\lambda = 2\pi(J/D)$. The Zeeman coupling term is described by $w_z$ where $\mathbf{H}$ is the applied field, $M_s$ is the saturation magnetization, and $\mu_0$ is the vacuum permeability. The magnetocrystalline anisotropy term is given by $w_a = -K_u(\mathbf{m} \cdot \hat{k})^2$ for uniaxial anisotropy and $w_a = -K_c(m_x^4 + m_y^4 + m_z^4)$ for cubic anisotropy, where $K_u$ and $\hat{k}$ are the uniaxial anisotropy strength and axis, and $K_c$ is the cubic anisotropy strength.

By scaling lengths by $\lambda$, magnetic field and anisotropy strength by critical values $H_D = D^2/\mu_0 M_s J$ and $K_0 = \mu_0 M_s H_D$ at which the magnetization is completely polarized, Eq. S1 can be rewritten in dimensionless parameters

$$\frac{E}{J\lambda} = \int d^3\mathbf{r} \left[\frac{1}{2}(\nabla \mathbf{m})^2 + 2\pi \mathbf{m} \cdot (\nabla \times \mathbf{m}) - 4\pi^2 \mathbf{m} \cdot \widetilde{\mathbf{H}} + \widetilde{w}_a\right], \quad (S2)$$

where $\widetilde{w}_a = -4\pi^2 \widetilde{K}_u(\mathbf{m} \cdot \hat{k})^2$ for uniaxial anisotropy and $\widetilde{w}_a = -4\pi^2 \widetilde{K}_c(m_x^4 + m_y^4 + m_z^4)$ for cubic anisotropy. $\widetilde{\mathbf{H}} = \mathbf{H}/H_D$ and $\widetilde{K}_{u(c)} = K_{u(c)}/K_0$. Evidently, the remaining free parameters are the magnetic field and anisotropy and Eq. S2 can be applied generally to chiral magnetic solids.

Numerical modeling of the energy minimization of $\boldsymbol{m}(\boldsymbol{r})$ based on Eq. S2 is performed by a variational-method-based relaxation routine [3–5]. The Hamiltonian is minimized iteratively by updating $\boldsymbol{m}(\boldsymbol{r})$ based on an update formula derived from the Euler-Lagrange equation of the system,

$$m_i^{\text{new}} = m_i^{\text{old}} - \frac{\text{MSTS}}{2}[E]_{m_i} \quad \text{(S3)}$$

where the subscript $i$ denotes spatial coordinates, $[E]_{m_i}$ denotes the functional derivative of $E$ with respect to $m_i$. MSTS is the maximum stable time step in the minimization routine, determined by the values of material parameters and the spacing of the computational grid [3–5]. $\boldsymbol{m}(\boldsymbol{r})$ is normalized after each iteration. The system is implied to be in a state corresponding to an energy minimum when the energy difference between iterations approaches zero. Iteratively, energy-minimizing $\boldsymbol{m}(\boldsymbol{r})$ and evolution dynamics can be derived. The immaterial helical wavevector field $\boldsymbol{q}(\boldsymbol{r})$ is derived from $\boldsymbol{m}(\boldsymbol{r})$ based on the eigenvector of the chirality tensor [6].

The 3D spatial discretization is performed on large 3D square-periodic grids with 24 grid points per helical wavelength $\lambda$. The spatial derivatives are calculated by finite difference methods with second-order accuracies based on central difference for the bulk nodes and the same scheme is used for the boundary nodes with imposed periodic boundary conditions. The dimensions of the computational volume are 6 to 11 helical wavelengths, in order to simulate heliknotons embedded in the bulk. The heliknoton structures in liquid crystals are used as initial conditions [6]. At finite applied fields and anisotropies, the helicoidal modulation periods deviate from the helical wavelength $\lambda = 2\pi(J/D)$. The dimensions of the computational volume are adjusted accordingly to avoid artificially induced frustration. The modulation period is determined by a Euler-type equation after solving the differential equation for a nonlinear pendulum [7]

$$\lambda' = \int_0^\pi \frac{2d\theta}{\sqrt{c+f(\theta)}}, \quad (S4)$$

where $f(\theta)$ is given by

$$f(\theta) = \begin{cases} -8\pi^2 \widetilde{H} \cos\theta & \text{(uniaxial easy} - \text{plane anisotropy, Fig. 3(a))} \\ -8\pi^2 [\widetilde{H} \cos\theta + \widetilde{K}_u \cos^2\theta] & \text{(uniaxial easy} - \text{axis anisotropy, Fig. 3(b))} \\ -8\pi^2 [\widetilde{H} \cos\theta + \widetilde{K}_c (\cos^4\theta + \sin^4\theta)]. & \text{(cubic anisotropy, Fig. 3(c))} \end{cases} \quad (S5)$$

The constant $c$ is the solution to the following integral equation

$$\int_0^\pi d\theta \sqrt{c+f(\theta)} = 2\pi^2. \quad (S6)$$

The field topology of energy-minimizing $\boldsymbol{m}(\boldsymbol{r})$ is determined by the $\pi_3(\mathbb{S}^2)$ topology charge – Hopf index. The Hopf index of a 3D topological soliton can be geometrically derived as the linking number of each pair of distinct preimages [8], as well as obtained via numerical integration as described in Ref. [9].

Fresnel mode images of Lorentz transmission electron microscopy are simulated based on diffraction of electron waves through magnetic materials. The equation describing the intensity of an image at small defocus $\Delta$ for a thin film normal to the electron beam direction $\hat{\boldsymbol{l}}$ is given by [10]

$$I(\boldsymbol{r}) = 1 - \int_0^d dl\, \Delta \frac{e\mu_0 \lambda_e}{2\pi\hbar} (\nabla \times \boldsymbol{m}(\boldsymbol{r})) \cdot \hat{\boldsymbol{l}} \quad (S7)$$

where $\lambda_e$ is the electron wavelength, $d$ is the film thickness, and $e$, $\mu_0$ and $\hbar$ are the electron charge, vacuum permeability, and the reduced Planck constant, respectively. The contrast of each image is then normalized.

## Supplementary References: